\documentclass[11pt,twocolumn,oneside]{article}
\usepackage{geometry}                % See geometry.pdf to learn the layout options. There are lots.
\usepackage{newtxtext,newtxmath}
\usepackage{graphicx}
\usepackage{amssymb,amsmath}
\usepackage{epstopdf}
\usepackage{authblk}
\usepackage{booktabs}
\usepackage[dvipsnames]{xcolor}
\usepackage{titlesec}
\usepackage{indentfirst}
\usepackage[labelfont=bf,font=small]{caption}

\titleformat*{\section}{\centering\large\bfseries}
\titleformat*{\subsection}{\normalsize\bfseries}
\titleformat*{\subsubsection}{\normalsize\bfseries}
\titleformat*{\paragraph}{\normalsize\bfseries}
\titleformat*{\subparagraph}{\normalsize\bfseries}

\usepackage[apaciteclassic]{apacite}

\newcommand{\dBPR}{\ensuremath{\text{d}\mathit{BPR}}}
\newcommand{\dVEL}{\ensuremath{\text{d}\mathit{VEL}}}
\newcommand{\BPR}{\ensuremath{\mathit{BPR}}}
\newcommand{\VEL}{\ensuremath{\mathit{VEL}}}

\newcommand{\setmeter}[2]{\ensuremath{%
\vcenter{\offinterlineskip
\halign{\hfil##\hfil\cr
$\scriptstyle#1$\cr
\noalign{\vskip1pt}
$\scriptstyle#2$\cr}
}}%
}

\title{
\Large\textbf{A Computational Study of the Role of Tonal Tension \linebreak in Expressive Piano Performance}
}
\author{ Carlos Cancino-Chac\'on$^{1,2}$ and Maarten Grachten$^{2}$}
\affil{ \normalsize\textit{${^1}$Austrian Research Institute for Artificial Intelligence, Vienna, Austria} \\ 
\textit{$^2$Institute of Computational Perception, Johannes Kepler University Linz, Austria} \\
carlos.cancino@ofai.at}
\date{}                                           % Activate to display a given date or no date

\begin{document}
\maketitle
\thispagestyle{empty}

\begin{abstract}
\vskip-2ex
\small
%Musicians do not perform a piece of notated music as an exact mechanical rendition of what is written in the score, but play it expressively through variations in tempo, dynamics and timbre.
\noindent Expressive variations of tempo and dynamics are an important aspect of music performances, involving a variety of underlying factors.
Previous work has showed a relation between such expressive variations (in particular expressive tempo) and perceptual characteristics derived from the musical score, such as musical expectations, and perceived tension.
%In this work we study the role of two probabilistic measures of musical expectation based on the IDyOM model \cite{Pearce:2005th} and three measures of tonal tension proposed by \citeA{Herremans:2016vra} in the prediction of expressive performances of classical piano music.
In this work we use a computational approach to study the role of three measures of tonal tension proposed by \citeA{Herremans:2016vra} in the prediction of expressive performances of classical piano music.
These features capture tonal relationships of the music represented in Chew's \emph{spiral array model}, a three dimensional representation of pitch classes, chords and keys constructed in such a way that spatial proximity represents close tonal relationships.
We use non-linear sequential models (recurrent neural networks) to assess the contribution of these features to the prediction of expressive dynamics and expressive tempo using a dataset of Mozart piano sonatas performed by a professional concert pianist.
Experiments of models trained with and without tonal tension features show that tonal tension helps predict \emph{change} of tempo and dynamics more than absolute tempo and dynamics values.
Furthermore, the improvement is stronger for dynamics than for tempo.
% A preliminary cross validation experiment suggests that the non-linear sequential models do not consistently benefit from tonal tension features when low-level pitch information is present.
% However, we do observe a small improvement when combining pitch, metrical and tonal tension features for predicting expressive parameters that describe changes in dynamics and tempo, when compared to models considering only pitch and metrical features.

%However, we do observe a substantial synergy of tonal tension features and metrical features, especially for predicting expressive parameters that describe dynamics.
% A preliminary  cross validation experiment suggests that tonal tension features might be more relevant for predicting local changes of tempo and loudness than for predicting the absolute values of these parameters.
% Furthermore, our results show that the internal representations of the score learned by recurrent networks trained for predicting expressive performances from low-level pitch information are implicitly related to tonal tension features.

% that the internal representations learned by the recurrent networks are related to tonal tension.
\end{abstract}

\section*{ Introduction}
\vskip-2ex
Expressive performance of music constitutes an important part of our enjoyment of several kinds of music, including Western art music and jazz.
In these kinds of music, an expressive performance is not expected to be an exact mechanical rendition of what is written in the score, but a combination of the performer's interpretation of both the intentions of the composer and its own expressive intentions that are conveyed to the listener through variations in dimensions such as tempo, dynamics and timbre.
Previous work has showed a relation between such expressive variations and perceptual characteristics derived from the musical score, such as musical expectations, and perceived tension \cite{Chew:2016ty,Farbood:2012by,Gingras:2016bm}.

%\cc{What is tension, what is tonal tension and what are tension-related features relevant for predicting expressive performance. }
%In this work, we use a data-driven computational model of expressive performance to estimate the contribution of tonal tension to the prediction of expressive tempo and dynamics.
The concept of musical tension is highly complex and multidimensional, and thus, difficult to formalize or quantify~\cite{Farbood:2012by,Herremans:2016vra}.
%According to \citeA{Farbood:2012by}, \emph{``[t]he phenomenon of tension is evident to listeners and is relatively easy to define in informal, qualitative terms; for example, increasing tension can be described as a feeling of rising intensity or impending climax, while decreasing tension can be described as a feeling of relaxation or resolution''}.
Informally,  \emph{``increasing tension can be described as a feeling of rising intensity or impending climax, while decreasing tension can be described as a feeling of relaxation or resolution''} \cite[pp.~387]{Farbood:2012by}.
The music cognition literature has shown that aspects related to musical tension include both psychological factors such as expectation and emotion; and musical factors such as rhythm, timing and dynamics and tonality.
%literature has shown that several aspects including psychological factors such as expectation and emotion, and musical factors such as rhythm, timing and dynamics and tonality.
%One of these many aspects is tonality [cite Lehrdal].
For a more thorough description of aspects that contribute to musical tension, we refer the reader to \cite{Farbood:2012by} and references therein.

\vskip-0.5ex
In this work we use a computational approach to study the role of tonal  tension features --as proposed by \cite{Herremans:2016vra}-- in the prediction of expressive performances of Classical piano music.
%Considering tonal tension as a feature for computational
Computational models of musical expression can be used to explain the way certain properties of a musical score relate to an expressive rendering of the music~{\cite{Widmer:2004bh}.
%Thus, these models provide
The KTH model \cite{Friberg:2006hs}, one of the most important \emph{rule-based} models of expressive performance includes rules that take into account tonal tension.
%\cc{Discuss related work in more detail and computational models of performance:
%KTH melodic charge: emphasize of notes relatively to the current chord and harmonic tension: emphasize chords that are far away relatively to the key}
%In this work we study the role of  tonal tension features --using a computational model proposed by \citeA{Herremans:2016vra}-- in the prediction of expressive performances of classical piano music.

%In this work we study the role of these two types of features, namely tonal tension features --using a model proposed by \citeA{Herremans:2016vra}-- and information theoretic expectancy features --using a probabilistic model of auditory expectation that computes information-theoretic features relating to the prediction of future events proposed by \citeA{Pearce:2005th}.

\vskip-0.5ex
The rest of this paper is structured as follows:
the Method section is divided into three subsections, the first of which describes the tonal tension and score features, followed by a brief description of the way expressive tempo and dynamics are quantified in this work; and finally, the recurrent neural network model relating the features to the expressive tempo and dynamics is presented.
The Experiments section describes the evaluation of the methods using cross-validation experiments.
Afterwards we discuss the results of these experiments and, finally, we present conclusions and future research directions.

\section*{ Method}\label{sec:method}
\vskip-2ex
This section details the computational methodology we use in this study.
It describes the features used to represent musical contexts -- the inputs of the model, the expressive parameters used to represent tempo and dynamics -- the outputs of the model, and the model itself.

\subsection*{Features}

\subsubsection*{Tonal Tension Features (T)}

In order to characterize tonal tension, we use a set of three quantities, which are computed using the method proposed by \citeA{Herremans:2016vra}.
These features capture tonal relationships of the music represented in Chew's \citeyear{Chew:2000wc} \emph{spiral array model}, a three dimensional representation of pitch classes, chords and keys constructed in such a way that spatial proximity represents close tonal relationships.
The tonal tension features are:
\begin{enumerate}
\item \emph{cloud diameter} ($T_{cd}$), which estimates the maximal tonal distance between notes in a segment of music;
\item \emph{cloud momentum} ($T_{cm}$) quantifies harmonic movement as the tonal distance from a section to the next; and
\item \emph{tensile strain} ($T_{ts}$), the relative tonal distance between the current segment and the center of effect of the key of the piece, i.e. the point in the spiral array mode that best represents the key of the piece.
\end{enumerate}
Since distances in the spiral array are can be large (in this particular work, an order of magnitude larger than the score features defined below), we scale the tension features described above by dividing them by the distance between enharmonically equivalent  notes (e.g.~C$\sharp$ and D$\flat$).
\citeA{Herremans:2016vra} evaluated these features by comparing them to the empirical study by \citeA{Farbood:2012by}, showing that these features correlate to human perception of tonal tension.

\subsubsection*{Score Features}\label{sec:score_features}
Following \citeA{CancinoChacon:2017uh}, we include two groups of low-level descriptors of a musical score that have been shown to predict characteristics of expressive performance.
These features provide a baseline showing to what degree expressive variations can be explained only by the nominal information in the score.

\begin{enumerate}
\item \textbf{Pitch (P)}
\begin{enumerate}
\item $(pitch_h, pitch_l, pitch_m)$.
Three features representing the chromatic pitch (as MIDI note numbers divided by $127$) of the highest note, the lowest note, and the melody note (if given, and zero otherwise) at each score position.
\item $(vic_1, vic_2, vic_3)$.\label{sec:pitch_vintcc}
Three features describing up to three vertical interval classes above the bass, i.e.~the intervals between the notes of a chord and the lowest pitch, excluding pitch class repetition and octaves.
For example, a $C$ major triad ($C$, $E$, $G$), starting at $C_4$ would be represented as $(\begin{array}{cccc}pitch_l &vic_1 &vic_2 &vic_3\end{array}) =  (\begin{array}{cccc}\frac{60}{127}& \frac{4}{11}& \frac{7}{11}&0\end{array})$, where $0$ denotes the absence of a third interval above $C_4$, i.e.~the absence of a fourth note in the chord.
\end{enumerate}

\item \textbf{Metrical (M)}
\begin{enumerate}
\item $b_{\phi,t}$.
The relative location of an onset within the bar,  computed as $b_{\phi, t} = \frac{t \mod B}{B}$, where $t$ is the temporal position of the onset measured in beats from the beginning of the score, and $B$ is the length of the bar in beats.
\item $(b_d, b_s, b_w)$.
Three binary features (taking values in $\{0, 1\}$) encoding the metrical strength of the $t$-th onset.
$b_d$ is nonzero at the downbeat (i.e.~whenever $b_{\phi, t} = 0$);
$b_s$ is nonzero at the secondary strong beat in duple meters (e.g.
quarter-note 3 in \setmeter{4}{4}, and eighth-note 4 in \setmeter{6}{8}), and $b_w$ is nonzero at weak metrical positions (i.e.~whenever $b_d$ and $b_s$ are both zero).
\end{enumerate}
\end{enumerate}

\subsection*{Expressive Parameters}
We consider an \emph{expressive parameter} to be a numerical descriptor that corresponds to common concepts involved in expressive performance.
In this section we briefly describe the parameters used to represent expressive tempo and dynamics.
%We represent an expressive parameter as $\vect{y}\in \mathbb{R}^T$, where $y_t$ represents the value of that parameter at onset $t$ in the score.

\begin{enumerate}
\item \textbf{Tempo}.
\begin{enumerate}
\item $\mathit{BPR}$.
We take the local \emph{beat period ratio} as a proxy for musical tempo.
%This parameter is computed as follows.
In order to compute this parameter, we average the performed onset times of all notes occurring at the same score position and then compute the $\mathit{BPR}$ by taking the slope of the averaged onset times (in seconds) with respect to the score onsets (in beats) and dividing the resulting series by its average beat period.
%We average the performed onset times (in seconds) of all notes occurring at the same score onset and then compute the $n$-th beat period as
%\begin{equation}
%\mathit{BP_n} = \frac{\text{IOI}_{perf_n}}{\text{IOI}_{score_n}},
%\end{equation}
%where $\text{IOI}_{perf_n}$ and $\text{IOI}_{score}$ are the inter-onset interval (IOI) in seconds of the performed onset times and the IOI in beats in the score, respectively.
%Then, we compute the $n$-th $\mathit{BPR}$ as
%\begin{equation}
%\mathit{BPR_n} = \frac{N \cdot \mathit{BP_n}}{\sum_{j=0}^{N-1}\mathit{BP_n}},
%\end{equation}
%where $N$ is the total number ofi.e. as the $n$-th beat period divided by the average beat period.
\item $\text{d}\mathit{BPR}$.
This parameter is computed as the first derivative of $\mathit{BPR}$ with respect to the score position, and corresponds to the  \emph{relative acceleration}, i.e.~the relative changes in musical tempo.
%In order to explore how well the features describe the \emph{relative} changes in tempo
%We average the performed onset times of all notes occurring at the same score onset and then compute the $\mathit{BPR}$ by taking the numerical derivative of the averaged onset times and dividing the resulting series by its average beat period.
\end{enumerate}
\item \textbf{Dynamics}.

\begin{enumerate}
\item $\mathit{VEL}$.
We treat the performed MIDI velocity as a proxy for the loudness of the note.
This parameter computed by taking the maximal performed MIDI velocity per score position, divided by $127$.
We use the terms loudness and dynamics interchangeably to refer to this parameter.
%\begin{equation}
%\mathit{VEL}_n = \frac{\max \text{VELOCITY}_{perf_n}}{127},
%\end{equation}
%where $\text{VELOCITY}_{perf_n}$ represents the performed MIDI velocity of all notes occurring at the $n$-th score onset.

\item $\text{d}\mathit{VEL}$.
This parameter is computed as the first derivative of $\mathit{VEL}$ with respect to the score position, and corresponds to the relative changes in loudness from one time step to the next.
\end{enumerate}

\end{enumerate}

\subsection*{Model}
We use recurrent neural networks (RNNs), a family of non-linear sequential models, to assess the contribution of the features described above to the prediction of expressive dynamics and tempo.
RNNs are a state-of-the-art family of neural architectures for modeling sequential data~\cite{Goodfellow-et-al-2016}.
These models have been used to model expressive dynamics and tempo~\cite{CancinoChacon:2017ht,Grachten:2017ub}.
In this work, we use a simple architecture, which we will refer to as bRNN, consisting of a composite bidirectional long short-term memory layer (LSTM) with multiplicative integration~\cite{Wu:2016vm} with 10 units (5 units processing information forwards and 5 processing information backwards) and a linear dense layer with a single unit as output.

\section*{ Experiments}
\vskip-2ex

In order to evaluate the contribution of the features described above to the prediction of expressive tempo and dynamics, we perform a cross-validation experiment to test the predictive quality of the model.
For this study, we use the Batik/Mozart dataset \cite{widmer02playingmozart}, which consists of recordings of 13 piano sonatas by W.~A.~Mozart performed by Austrian concert pianist Roland Batik which have been aligned to their scores.
An important characteristic of this dataset is that the melody voices are manually identified.
These performances were recorded on a B\"osendorfer SE 290, a computer controlled grand piano.

%We perform a 5-fold cross validation experiment for each model trained on different feature sets for each expressive parameter.
For each expressive parameter, we perform eight 5-fold cross-validation experiments corresponding  to models trained on  all combinations of feature sets, i.e.
all combinations of pitch features (\textbf{P}), metrical features (\textbf{M}) and tension features (\textbf{T});
as well as a feature set consisting of a selection of features using a univariate feature selection method (\textbf{FS}).
Each 5-fold cross-validation experiment is conducted as follows:
each model is trained/tested on 5 different partitions (folds) of the dataset, which is organized into training and test sets, such that each piece in the corpus occurs exactly one in the test set.
For each fold, we use 80\% of the pieces for training and 20\% for testing the model.
The parameters of the model are learned by minimizing the mean squared error on the training set using RMSProp, a variant of the stochastic gradient descent algorithm~\cite{Tieleman:2012kl}.

The feature selection procedure computes the pairwise \emph{mutual information} between each of the features and each of the expressive parameters.
This information-theoretic measure expresses how much knowing the value of one variable reduces uncertainty about the value of the other variable~\cite{Ross:2014kl}, and is a common way of determining the relevance of features in prediction tasks.
% , between each individual feature and the expressive parameters\footnote{
% We compute estimate the mutual information between features and expressive parameters using the implementation in the scikit-learn python library, which is based on the method proposed in \cite{Ross:2014kl}.}.
Formally, the mutual information between two variables $x$ and $y$ is given by
\begin{equation*}
\text{MI}(x, y) = \mathbb{E}\left\{\log\left(\frac{p(x, y)}{p(x) p(y)} \right)\right\},
\end{equation*}
where $\mathbb{E}$ is the expectation operator, $p(x,y)$ is the joint probability distribution of $x$ and $y$, and $p(x),p(y)$ are the marginal probability distributions of $x$ and $y$, respectively.
If $x$ and $y$ are statistically independent (i.e.~they do not share information about each other), the mutual information is zero.
In the \textbf{FS} scenario we select the 10 features with the largest mutual information for each expressive parameter.
This procedure was performed on a small subset of the Batik/Mozart dataset (20\% of the pieces selected randomly).

%Additionally, we use a univariate feature selection method (mutual information) to select the top most relevant features according to the mutual information [Describe method].
%This feature set is denoted as \textbf{FS}.
%Each of these cross validation experiments is conducted as follows:

%the dataset is divided into 5 partitions (folds), and then each model is trained on four of these partitions (using 80\% of the pieces for updating the model parameters and 20\% for validation) and then tested on the remaining partition.
%
%%For each fold, we use 80\% of the pieces for training and 20\% for validating the model.
%The parameters of the model are learned by minimizing the mean squared error ($MSE$) using a variant of stochastic gradient descent.
%For reproducibility, the hyper-parameters were selected as follows:
%The learning rate for RMS-prop is set to $1\times10^{-3}$;
%The $l_2$-norm weight regularization coefficient is set to $1\times 10^{-3}$;
%The gradients are clipped at $2$,
%The weights of the LSTM layers were initialized from $\mathcal{U}(-0.01, 0.01)$.
%The weights of the output layer were initialized using the Glorot initialization~\cite{Glorot:2010uc}.

\section*{ Discussion}
\vskip-2ex
Figure \ref{fig:feature_selection} shows the mutual information between each feature and the expressive parameters, normalized for each expressive parameter.
In this plot, the height of a column (the value of the mutual information) signifies how closely related that feature is to the expressive parameter.
%Although the results in Table \ref{tab:5fcv_results} for the feature set consisting of the 10 features with the largest mutual information between the features and targets (\textbf{FS}) are not the best for each target, the difference between the predictive accuracy of the models trained on the feature selection and the best results is not statistically significant, according to Tukey's HSD test.
The results in this plot suggest that the tension features, in particular the cloud diameter might be more related to the prediction of changes in tempo and dynamics, whereas the tensile strain might be more related to absolute tempo and dynamics than their changes.
From a musical perspective, these findings seem plausible, since cloud diameter refers to melodic events, whose performance might be dependent on the character of the passage, whereas tensile strain depends on structural harmonic characteristics of the music.
%Although the results using the feature selection method were not the best, the HSD test reveals that the difference between these results and the best are not statistically significant.
%The MI method does not consider the temporal dependencies, which according to Figure \ref{fig:sensitivity} are relevant.

\begin{figure}[t]
\centering
\includegraphics[width=\linewidth]{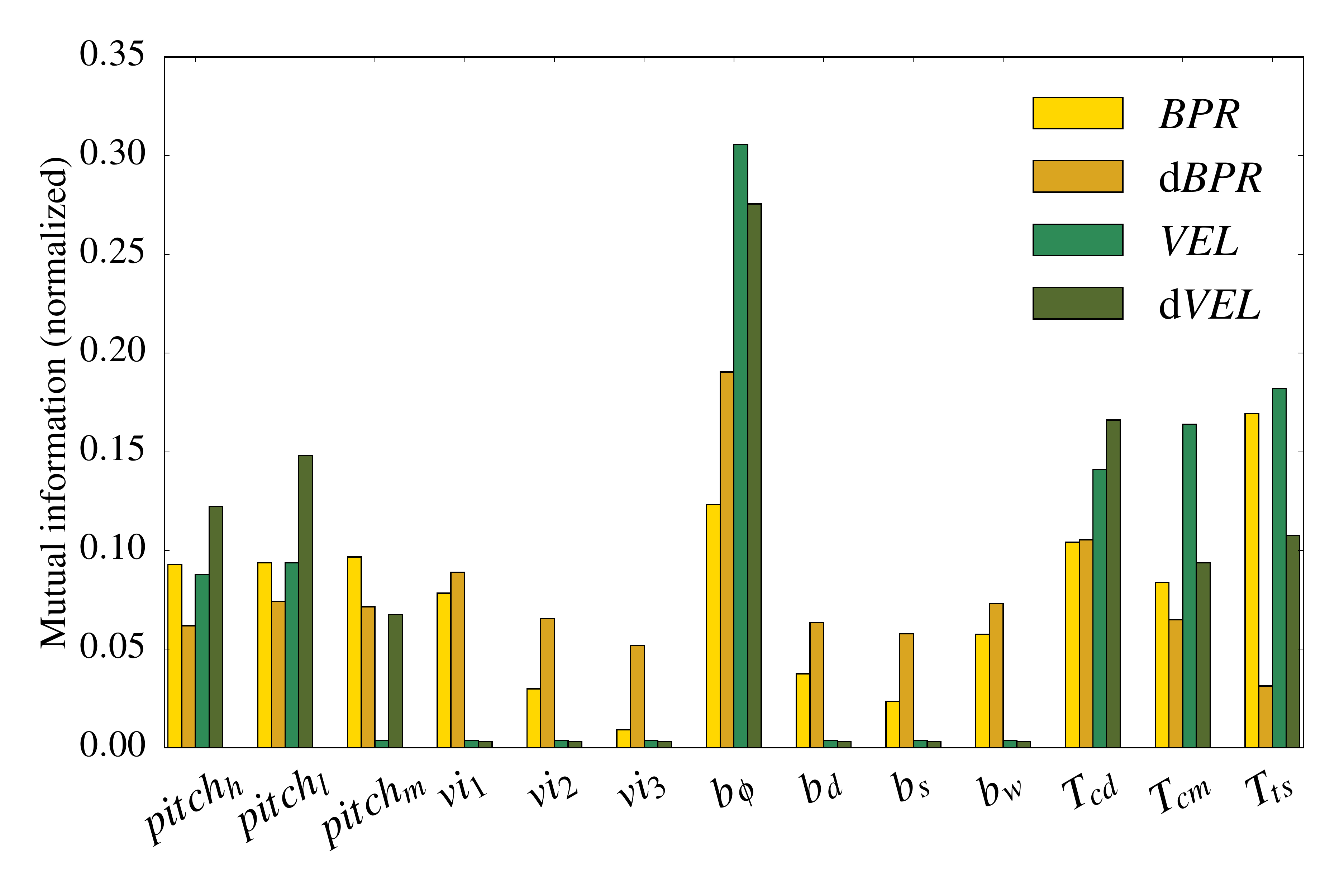}
\caption{\small\textbf{Normalized mutual information between features and expressive parameters.
Larger mutual information means that the variables are more related to each other.}}
\label{fig:feature_selection}
\end{figure}

Note that the values in Figure~\ref{fig:feature_selection} only measure the MI of the features and expressive parameters at isolated time instances, without context.
Although this gives a good first impression of the relevance of features, the bRNN model presented above is specifically designed to take advantage of the temporal context to make predictions, implying that feature values at times before and after $\tau$ may also help to predict expressive parameters at $\tau$.
Therefore Figure \ref{fig:feature_selection} does not necessarily reflect the relevance of features when used as input to the bRNN.

Table \ref{tab:5fcv_results} displays the results of the cross-validation experiments using the bRNN on different feature sets.
We evaluate model accuracy with the coefficient of determination ($R^2$).
This measure expresses the proportion of variance of the expressive parameters that can be explained by the models as a function of the feature set. % and Pearson's correlation coefficient $r$.
% To examine the differences between the $R^2$ values of all feature sets in Table \ref{tab:5fcv_results}, we performed a separate one-way ANOVA for each expressive parameter ($\mathit{BPR}$, $\text{d}\mathit{BPR}$, $\mathit{VEL}$ and $\text{d}\mathit{VEL}$).
% \mg{Is this across all feature combinations? I don't really get what this anova measures.}
% These differences were statistically significant in all cases at the $p<0.01$ level as measured by Fisher's $F$ ratio.
%To examine the differences between the $R^2$ values of all feature sets in Table \ref{tab:5fcv_results}, we performed a separate one-way ANOVA across all feature combinations for each expressive parameter ($\mathit{BPR}$, $\text{d}\mathit{BPR}$, $\mathit{VEL}$ and $\text{d}\mathit{VEL}$).
%This yielded a significant effect of feature set in all cases at the $p<0.05$ level as measured by Fisher's $F$ ratio.
%Post-hoc Tukey's HSD tests adjusted for multiple comparisons indicated some significant differences, which we show in Table \ref{tab:5fcv_results} as bold vs.~non-bold numbers.
For each expressive parameter, we conducted a paired--samples two-tailed t-test at the $p<0.01$ level to compare the differences between the $R^2$ values of features sets with  and without tension features
(i.e. \textbf{P} vs.~\textbf{P} + \textbf{T},\ \ \textbf{M} vs.~\textbf{M} + \textbf{T},\ \  and \textbf{M} + \textbf{P} vs.~\textbf{M} + \textbf{P} + \textbf{T}).
The effect of including \textbf{T} was not significant for the prediction of \BPR\ and \VEL\, but it \emph{was} significant for the prediction of \dVEL\ in all cases.
For \dBPR\ including \textbf{T} was only beneficial in combination with \textbf{P}.
Based on this result we hypothesize that the concept of tonal tension is relevant for changes in tempo, but the features used to represent tonal tension may not convey enough information by themselves and are therefore only advantageous in combination with more specific pitch information.

\begin{table}[t]
\centering

% \caption{\small \textbf{Proportion of variance explained ($R^2$) for expressive tempo and dynamics using different feature sets, averaged over all pieces on the Batik/Mozart corpus.
%   Larger $R^2$ values mean more accurate predictions. 
%  Bold entries are those feature set combinations + \textbf{T} that differ significantly from the combination without \textbf{T}.}}
\caption{\small \textbf{Proportion of variance explained ($R^2$) for expressive tempo and dynamics using different feature sets, averaged over all pieces on the Batik/Mozart corpus.
    Larger $R^2$ values reflect more accurate predictions.
    For each combination of target and feature set, the results are listed for that feature set as is (left), and including tonal tension features \textbf{T} (right).
    For clarity, improvements of $R^2$ as a result of adding \textbf{T} are marked in green, whereas detriments are marked in red.
    Bold numbers mark a statistically significant difference ($p < 0.01$).
    The effect size (Cohen's $d$) is reported in parenthesis for those cases with statistically significant differences.
  }}
\label{tab:5fcv_results}
 \newcommand{\cwa}{\hspace{1em}}
 \newcommand{\cwb}{\hspace{2.5em}}
 \newcommand{\cwc}{\hspace{3.5em}}

\newcommand{\sw}{\hskip 1.5em}
\newcommand{\wrs}{\color{Black!10!Red!90!}}
\newcommand{\btr}{\color{Black!10!Green!90!}}

% \begin{tabular}{cl@{\cwa}l@{\cwb}l@{\cwa}l@{\cwc}l@{\cwa}l@{\cwb}l@{\cwa}l}
%\begin{tabular}{@{}cccccc@{\cwc}cccc@{}}
%\toprule
%                        & \multicolumn{4}{c}{Tempo}                &            & \multicolumn{4}{c}{Dynamics}                        \\
%                        & \multicolumn{2}{c}{\BPR} & \multicolumn{2}{c}{\dBPR} & &\multicolumn{2}{c}{\VEL} & \multicolumn{2}{c}{\dVEL} \\
%Feature Set             & -       & + \textbf{T}   & -        & + \textbf{T}   & & -      & + \textbf{T}   & -      & + \textbf{T}    \\ \midrule
%$\emptyset$             & -       & 0.010          & -        & 0.011 &          & -       & 0.018          & -      & 0.026           \\
%\textbf{P}              & 0.024   & 0.021          & 0.068    & 0.073 &         & 0.326   & 0.335          & 0.236  & 0.250           \\
%\textbf{M}              & 0.051   & 0.054          & 0.093    & 0.092 &         & 0.048   & 0.052          & 0.041  & 0.050           \\
%\textbf{P} + \textbf{M} & 0.056   & 0.054          & 0.105    & 0.110 &         & 0.351   & 0.347          & 0.250  & \textbf{0.282}  \\ \bottomrule

\begin{tabular}{@{}c@{\sw}cc@{\sw}ccc@{}}
\toprule
                        & \multicolumn{4}{c}{Tempo}                     \\
                        & \multicolumn{2}{c@{\sw}}{\BPR} & \multicolumn{2}{c}{\dBPR}  \\
  \cmidrule(r{1em}){2-3} \cmidrule{4-6}
Feature Set             &         & + \textbf{T}   &          & + \textbf{T}     & $d$\\ \midrule
$\emptyset$             & -      &      0.010  & -     &      0.011 \\
\textbf{P}              & 0.024  & \wrs 0.021  & 0.068 & \btr \textbf{0.073} &(0.10)\\
\textbf{M}              & 0.051  & \btr 0.054  & 0.093 & \wrs 0.092 \\
\textbf{P} + \textbf{M} & 0.056  & \wrs 0.054  & 0.105 & \btr \textbf{0.110} &(0.06)\\ \midrule
% $\emptyset$             & -       & 0.010          & -        & 0.011           \\
% \textbf{P}              & 0.024   & 0.021          & 0.068    & 0.073           \\
% \textbf{M}              & 0.051   & 0.054          & 0.093    & 0.092            \\
% \textbf{P} + \textbf{M} & 0.056   & 0.054          & 0.105    & 0.110  \\ \midrule
& \multicolumn{4}{c}{Dynamics}                        \\
&\multicolumn{2}{c@{\sw}}{\VEL} & \multicolumn{2}{c}{\dVEL} \\
  \cmidrule(r{1em}){2-3} \cmidrule{4-6}
Feature Set             &        & + \textbf{T}   &         & + \textbf{T}  & $d$ \\ \midrule
$\emptyset$             & -     &      0.018 & -      & 0.026                \\            
\textbf{P}              & 0.326 & \btr 0.335 & 0.236  & \btr \textbf{0.250}  &(0.16)         \\            
\textbf{M}              & 0.048 & \btr 0.052 & 0.041  & \btr \textbf{0.050} &(0.18)          \\            
\textbf{P} + \textbf{M} & 0.351 & \wrs 0.347 & 0.250  & \btr \textbf{0.282} &(0.40)  \\ \bottomrule
% $\emptyset$             & -       & 0.018          & -      & 0.026           \\
% \textbf{P}              & 0.326   & 0.335          & 0.236  & 0.250           \\
% \textbf{M}              & 0.048   & 0.052          & 0.041  & 0.050           \\
% \textbf{P} + \textbf{M} & 0.351   & 0.347          & 0.250  & \textbf{0.282}  \\ \bottomrule

% \caption{\small \textbf{Proportion of variance explained ($R^2$) for expressive tempo and dynamics using different feature setsn, averaged over all pieces on the Batik/Mozart corpus.
%     Larger $R^2$ values mean more accurate predictions.
%     For each combination of target and feature set, the results are listed for that feature set excluding and including tonal tension features \textbf{T}.
%     For clarity, improvements of $R^2$ as a result of adding \textbf{T} are marked in green, whereas detriments are marked in red.
%     Bold numbers mark a statistically significant difference.
%   }}
% \label{tab:5fcv_results}
% \newcommand{\sw}{\hskip 3em}
% \newcommand{\wrs}{\color{Black!10!Red!90!}}
% \newcommand{\btr}{\color{Black!10!Green!90!}}

% \begin{tabular}{c@{\sw}cc@{\sw}cc@{\sw}@{\sw}cc@{\sw}cc}
% \toprule
%                         & \multicolumn{4}{c@{\sw}@{\sw}}{Tempo}                            & \multicolumn{4}{c}{Dynamics}                        \\
%                         & \multicolumn{2}{c@{\sw}}{\BPR} & \multicolumn{2}{c@{\sw}@{\sw}}{\dBPR} & \multicolumn{2}{c@{\sw}}{\VEL} & \multicolumn{2}{c}{\dVEL} \\
% Feature Set             & -      & + \textbf{T}& -     &+ \textbf{T}& -     &+ \textbf{T}& -      & + \textbf{T}         \\ \midrule

\end{tabular}
\end{table}

\begin{figure*}[t]
\centering
\includegraphics[width=0.49\linewidth]{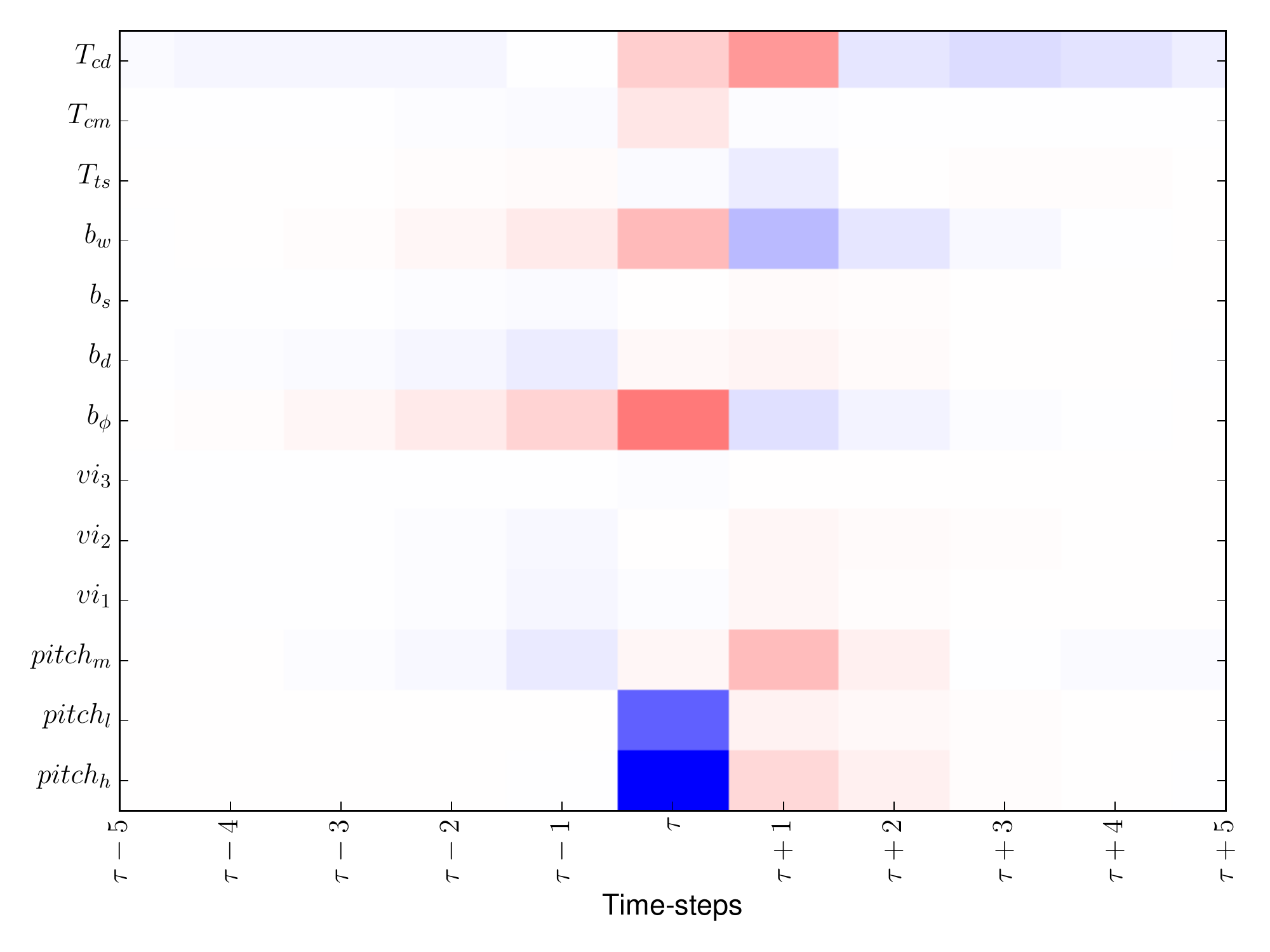}
\includegraphics[width=0.49\linewidth]{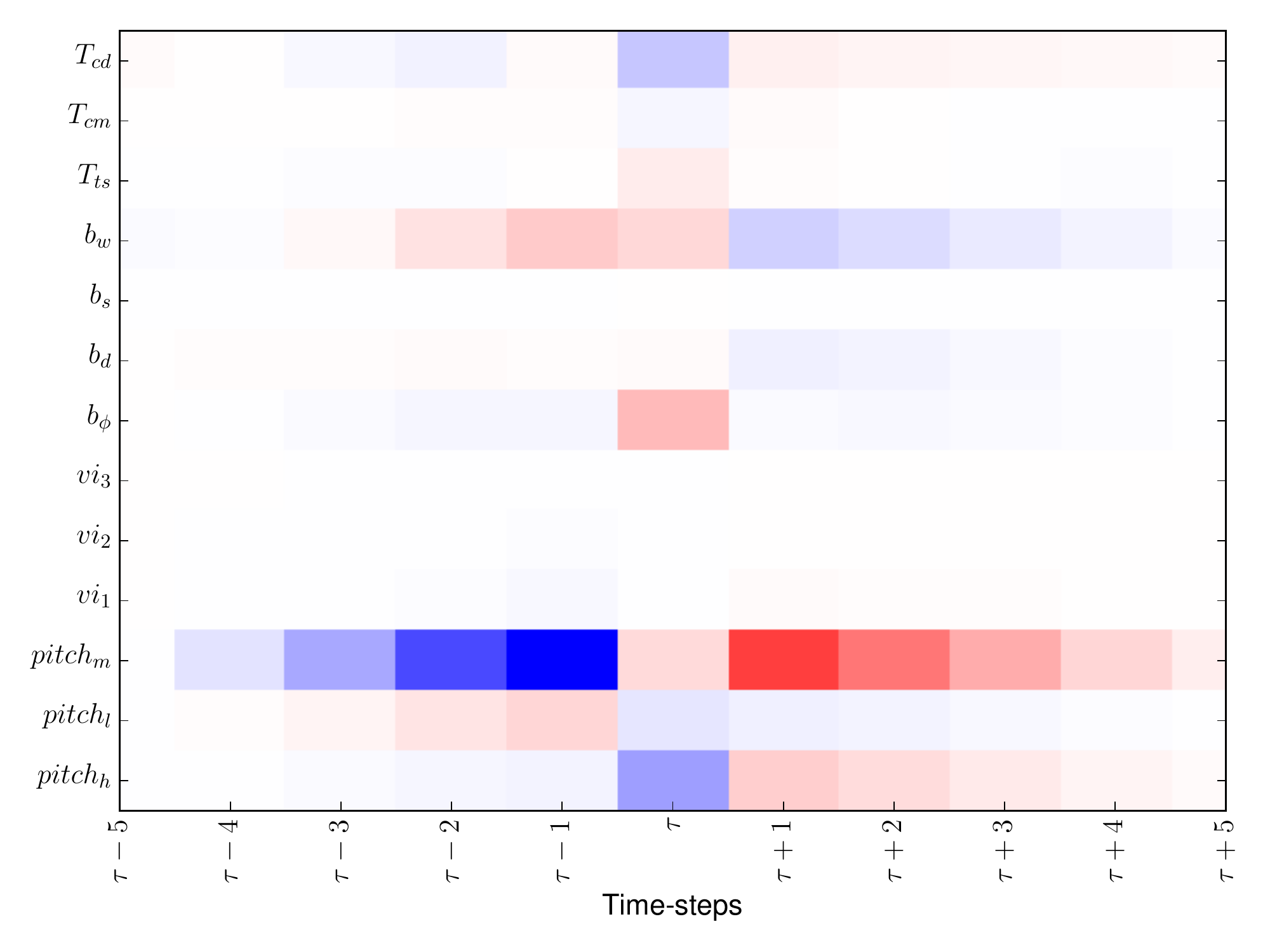}
\caption{\small \textbf{Differential sensitivity plots for} \dBPR{} \textbf{(left) and} \dVEL{} \textbf{(right).
Each row in the plot corresponds to an input feature and each column to the contribution of its value at that time-step to the output of the model at time $\tau$ (the center of each plot).
Red and blue indicate a positive (i.e.~slowing down or increasing loudness) and negative contribution (i.e.~speeding up or decreasing loudness), respectively.}}
\label{fig:sensitivity}
\end{figure*}

In order to visualize the contribution of each feature to the prediction of the changes in tempo and dynamics, we perform a differential sensitivity analysis\footnote{We use the definition of sensitivity analysis from the applied mathematics literature, not in the sense used in the psychology literature.} of the models by computing a local linear approximation of the output of the bRNNs trained on all features (\textbf{P} + \textbf{M} + \textbf{T}).
The resulting sensitivities are plotted in Figure~\ref{fig:sensitivity}. This figure can by roughly interpreted as follows: The color in the cell that corresponds to feature $f$ and time step $t$ represents the contribution of the value of $f$ at time $t$ to the prediction of the expressive parameter at time $t = \tau$ (the center column of the plot).
Blue tones reflect feature values that negatively contribute to the predicted value (the higher the feature value the lower the predicted value), and red tones reflect feature values that positively contribute to the predicted value.

%Although these plots show that certain score features play a more prominent role in predicting the expressive parameters, we will focus here in the contribution of the tonal tension features.
The plots follow a similar trend to the results showed in Figure~\ref{fig:feature_selection}, where the features with higher mutual information also have brighter colors in Figure~\ref{fig:sensitivity}, with the added benefit that the contribution of each feature at different time steps can be visualized.
%Also, the relatively strong sensitivity of the model to the temporal context of the pitch features (especially in the right plot) is compatible with the hypothesis that the model learns features that represent (aspects of) the tonal context directly from the pitch 

The plots in Figure~\ref{fig:sensitivity} suggest a tendency of the performer to emphasize melodic and harmonic events by slowing down.
For example, a chromatic melody note in an otherwise tonally stable section of the piece, (like the presentation of the main theme during the exposition of a sonata)  is emphasized by slowing down (see the reddish hue in $T_{cd}$ for time-steps $\tau$ and $\tau + 1$ in the left plot); while upcoming chromatic notes contribute to speeding up (the bluish hue in $T_{cd}$ for time-steps $> \tau + 1$).
%This finding is consistent with the \emph{melodic charge} rule in the KTH model \cite{Friberg:2006hs}.
On the other hand, chords that are tonally far from the current key are emphasized by slowing and an increase in loudness (the reddish hue for time-step $\tau$ in $T_{cm}$ in the left plot and in $T_{ts}$ in the right plot).
Furthermore, upcoming sections with modulations to distant keys (like the more unstable parts of the development of a sonata) contribute to speeding up (the bluish hue in $T_{cd}$ for time-steps $> \tau + 2$ in the left plot)
These findings agree with common performance rules (c.f.
the \emph{melodic charge} and \emph{harmonic charge} rules in the KTH model \cite{Friberg:2006hs}).

% A possible explanation for this could be the fact that  bRNNs might be powerful enough to learn implicit  representations of musical features related to high-level concepts such as tonal tension and musical structure directly from the low-level pitch features.
% \cc{In order to test whether the bRNN learns high level features related to tonal tension, we compare the mutual information between each pitch and hidden unit activations to the tension features.
% This comparison is showed in Figure~\ref{fig:mi_pitch_hidden}.
% In this plot we can see that the mutual information between the hidden activations and the tension features is higher, suggesting that the bRNN is implicitly learning representations of concepts related to tension.
% Furthermore, we can see that this is more relevant for the forward component, suggesting that processing of the pitch features in the way we experience music (in a causal way) might be more relevant.}

% \begin{figure*}[t]
% \centering
% %\includegraphics[width=0.75\linewidth]{mi_pitch_tension_dibi}
% \includegraphics[width=0.95\linewidth]{mutual_info_icmpc}
% \caption{\small Comparison of the mutual information between tonal tension features and pitch features and hidden layer activations.
% The plot on top represents the models trained for predicting $\textit{dIBI}$ and the bottom plot represents the models trained for predicting $\textit{dVEL}$.
% $h_{f_i}$ and $h_{b_i}$ correspond to the $i$-th unit in the forward and backward component of the LSTM layer.}
% \label{fig:mi_pitch_hidden}
% \end{figure*}

\section*{ Conclusions}
\vskip-2ex
In this work we have empirically investigated the role of tonal tension in shaping musical expression in classical piano performances.
Our experimental results show that using tonal tension information improves predictions of \emph{change} of tempo and dynamics, but not predictions of absolute tempo and dynamics.
For predicting changes in tempo, using tonal tension features as defined in \cite{Herremans:2016vra} was only beneficial when low level pitch information was also available.
This suggests that tonal tension features are potentially relevant for predicting tempo changes, but by themselves not sufficiently specific for that purpose.
% When regarded in a simple one-to-one relation, features describing tonal tension are related to expressive parameters in terms of mutual information.
% However, our experiments also show that in sufficiently powerful models such as the bi-directional recurrent networks used here, using low level pitch information makes tonal tension features superfluous for predicting expression.
% We hypothesize that the reason for this is the fact that the pitch information subsumes the tonal tension features, and that the models may be able to implicitly learn tonal characteristics from the pitch information to the degree that they are relevant for musical expression.

%\textbf{An interesting observation is that tonal tension features combined with pitch and metrical features improve predictions of loudness changes}
%A particularly interesting observation is that tonal tension features combined with metrical features allows for more accurate models than those with either tonal tension or metrical features alone.
%This synergetic effect was strongest for the expressive parameters describing dynamics.

Furthermore, an analysis of the trained models corroborates previously formulated relationships between performance and tension, as defined in the KTH model.

Future work may focus on a more explicit testing of the hypothesis that recurrent neural network models may learn features describing tonal characteristics from low level pitch information as a side effect of learning to predict expressive tempo and dynamics.

%However, this relatedness does not necessarily translate to more accurate predictive models when those tonal features are included as inputs to the model.
%This work presented a computational approach to quantify the contribution of tonal tension features to the prediction of expressive tempo and dynamics.

% Future work might involve joint models that consider other perceptually relevant features, such as the expectancy features~\cite{CancinoChacon:2017uh}.
% \cc{Furthermore, the results of the analysis of the internal representations of the neural network suggest that these models can learn representations that are related to perceptual features.
% Such approaches may be used to propose alternative characterizations of perceptual concepts such as musical tension and expectation.}
% %This study is expected to quantify the contribution of harmonic tension features to the prediction of expressive tempo and dynamics.
% The use of such perceptually features may lead to more meaningful computational models of expressive performance.

\paragraph*{Acknowledgments.}
Funding for this work was provided by the European Research Council (ERC) under the EU's Horizon 2020 Framework Programme (ERC Grant Agreement number 670035,
project ``Con Espressione'').
\newpage

\bibliographystyle{apacite}

\bibliography{bib_cc.bib}

\end{document}